\documentclass[11pt, a4paper]{article}
\usepackage[]{graphicx}
\usepackage[]{color}
\usepackage{xcolor}
\makeatletter
\def\maxwidth{ %
  \ifdim\Gin@nat@width>\linewidth
    \linewidth
  \else
    \Gin@nat@width
  \fi
}
\makeatother

\definecolor{fgcolor}{rgb}{0.345, 0.345, 0.345}
\newcommand{\hlnum}[1]{\textcolor[rgb]{0.686,0.059,0.569}{#1}}%
\newcommand{\hlstr}[1]{\textcolor[rgb]{0.192,0.494,0.8}{#1}}%
\newcommand{\hlopt}[1]{\textcolor[rgb]{0,0,0}{#1}}%
\newcommand{\hlstd}[1]{\textcolor[rgb]{0.345,0.345,0.345}{#1}}%
\newcommand{\hlkwb}[1]{\textcolor[rgb]{0.69,0.353,0.396}{#1}}%
\newcommand{\hlkwc}[1]{\textcolor[rgb]{0.333,0.667,0.333}{#1}}%
\newcommand{\hlkwd}[1]{\textcolor[rgb]{0.737,0.353,0.396}{\textbf{#1}}}%

\usepackage{framed}
\makeatletter
\newenvironment{kframe}{

 \def\at@end@of@kframe{}%
 \ifinner\ifhmode%
  \def\at@end@of@kframe{\end{minipage}}%
  \begin{minipage}{\columnwidth}%
 \fi\fi%
 \def\FrameCommand##1{\hskip\@totalleftmargin \hskip-\fboxsep
 \colorbox{shadecolor}{##1}\hskip-\fboxsep
     \hskip-\linewidth \hskip-\@totalleftmargin \hskip\columnwidth}%
 \MakeFramed {\advance\hsize-\width
   \@totalleftmargin\z@ \linewidth\hsize
   \@setminipage}}%
 {\par\unskip\endMakeFramed%
 \at@end@of@kframe}
\makeatother

\definecolor{shadecolor}{rgb}{.97, .97, .97}
\definecolor{messagecolor}{rgb}{0, 0, 0}
\definecolor{warningcolor}{rgb}{1, 0, 1}
\definecolor{errorcolor}{rgb}{1, 0, 0}
\newenvironment{knitrout}{}{} % an empty environment to be redefined in TeX

\usepackage{alltt}
\usepackage{makecell}
\usepackage{verbatim}
\usepackage[vmargin=3cm, hmargin=2cm]{geometry}
\usepackage{url}
\usepackage{fancyhdr}
\usepackage{amsmath, amssymb}
\usepackage{longtable}
\usepackage{lscape}

\usepackage{xspace}
\usepackage{booktabs}
\usepackage{newfloat}
\usepackage{rotating}
\usepackage{scrextend}
\usepackage{csquotes}
\DeclareFloatingEnvironment[
    placement=!htbp
]{listing}
\usepackage{fancyvrb}
\usepackage{lmodern}
\usepackage{nameref}
\usepackage[T1]{fontenc}
\usepackage{listings}
\usepackage[colorlinks=true]{hyperref}
\usepackage{bm}

\usepackage[font=sf, labelfont={sf}, margin=1cm]{caption}

\usepackage[latin1]{inputenc}

\usepackage{natbib}

\IfFileExists{upquote.sty}{\usepackage{upquote}}{}

\begin{document}

\hypersetup{
    colorlinks,
    linkcolor={red!50!black},
    citecolor={blue!50!black},
    urlcolor={blue!80!black}
}

\begin{center}
    \Large
    \textbf{Shrinkage estimation for dose-response modeling in phase II trials with multiple schedules}

    \vspace{0.4cm}
    \large
    \textbf{Burak K\"ursad G\"unhan},\footnote{\textit{Department of Medical Statistics, University Medical Center G\"ottingen, G\"ottingen, Germany} \label{goettingen}} \footnote{\textit{Correspondence to: Burak K\"ursad G\"unhan; email: \texttt{burak.gunhan@med.uni-goettingen.de}}}  \textbf{Paul Meyvisch},\footnote{\textit{Galapagos NV, Mechelen, Belgium}}  \textbf{Tim Friede}\footref{goettingen} 
    \vspace{0.9cm}
\end{center}

%%%%%%%%%%%%%%%%%%%%%%%%%%%%%%%%%%%%%%%%%%%%%%%%%%%%%%
% ABSTRACT
%%%%%%%%%%%%%%%%%%%%%%%%%%%%%%%%%%%%%%%%%%%%%%%%%%%%%%
Recently, phase II trials with multiple schedules (frequency of administrations) have become more popular, for instance in the development of treatments for atopic dermatitis. If the relationship of the dose and response is described by a parametric model, a simplistic approach is to pool doses from different schedules. However, this approach ignores the potential heterogeneity in dose-response curves between schedules. A more reasonable approach is the partial pooling, i.e. certain parameters of the dose-response curves are shared, while others are allowed to vary. Rather than using schedule-specific fixed-effects, we propose a Bayesian hierarchical model with random-effects to model the between-schedule heterogeneity with regard to certain parameters. Schedule-specific dose-response relationships can then be estimated using shrinkage estimation. Considering Emax models, the proposed method displayed desirable performance in terms of the mean absolute error and the coverage probabilities for the dose-response curve compared to the complete pooling. Furthermore, it outperformed the partial pooling with schedule-specific fixed-effects by producing lower mean absolute error and shorter credible intervals. The methods are illustrated using simulations and a phase II trial example in atopic dermatitis. A publicly available R package, \texttt{ModStan}, is developed to automate the implementation of the proposed method (\href{https://github.com/gunhanb/ModStan}{https://github.com/gunhanb/ModStan}).\\


%%%%%%%%%%%%%%%%%%%%%%%%%%%%%%%%%%%%%%%%%%%%%%%%%%%%%%
\section{Introduction}\label{intro}
%%%%%%%%%%%%%%%%%%%%%%%%%%%%%%%%%%%%%%%%%%%%%%%%%%%%%%
In phase II of any clinical development program, the investigations of the dose-response relationship of a compound is crucial. Usually, there are two main goals of these investigations: (a) establishing a dose-response signal and (b) estimating the dose-response function \citep{ruberg}. In addition to the dose, a treatment plan of a phase II trial includes the \emph{schedule} (or dose regimen), that is the frequency of the administration, for instance a weekly or biweekly schedule. Recently, phase II trials with multiple schedules have become more popular, for instance in the development of monoclonal antibodies as treatments for a variety of diseases including hypercholesterolaemia \citep{Giugliano2012} and atopic dermatitis \citep{dupilumab}. \cite{eichenfield2017} reviewed many therapies for atopic dermatitis which were in phase II or III of clinical development. Multiple schedules were investigated in phase II trials of almost half of the investigated therapies \citep{eichenfield2017}. However, standard methods for dose-response estimation cannot account for multiple schedules.

Estimating separate dose-response curves for each schedule by a parametric model is a one way to tackle this problem, that is full stratification of the dose-response curves. However, this method ignores the information shared between different schedules. Alternatively, one can completely pool doses from different schedules.
The main problem with complete pooling is that it does not take into account the potential heterogeneity between different schedules. A more reasonable approach is the \emph{partial pooling}, that is certain parameters of the dose-response curves are shared, while others are allowed to vary. A placebo effect parameter can be, reasonably, assumed shared between schedules, whereas this may not be true for the $\text{ED}_{50}$ parameter, the dose at which half of the maximum effect is reached. \cite{Feller2017} proposed a partial pooling approach in which unshared parameters are treated as schedule-specific fixed-effects \citep{Mollenhof2019}.

We consider trials with (very) few schedules of small to moderate size. Here, borrowing available information is of great interest. Rather than using schedule-specific fixed-effects, we propose a Bayesian hierarchical model with random-effects to model the between-schedule heterogeneity with regard to certain parameters. schedule-specific parameters can be estimated using \emph{shrinkage estimation}. The basic idea of the shrinkage estimation is that stratified parameter estimates can be improved by shrinking towards the population mean. It has been shown that the shrinkage estimation improves the estimation accuracy in comparison to estimates obtained by pooling or stratification \citep{Efron1975}. Shrinkage estimation in the context of clinical trials were investigated by \cite{Jones2011} and \cite{Freidlin1326} among others. A popular application is the estimation of the treatment effect in the presence of subgroups, for example estimating response rate in a phase II trial with multiple patient populations \citep{EXNEX}. Here, we are interested in parametric dose-response models in the presence of multiple schedules, hence shrinkage estimators of the parameters of a dose-response model, for example the $\text{ED}_{50}$ parameter of an Emax model. Shrinkage estimation allows \emph{dynamic borrowing} \citep{viele}, in which the weights for each schedule depend on the data instead of using fixed weights. Dynamic borrowing results in considerable gain in efficiency, while being a robust method against the heterogeneity between schedules. A theoretical justification for shrinkage can be established through the concept of \emph{exchangeability} of the parameters between schedules. This means that finding no systematic reason to distinguish schedule-specific parameters, in other words, they are similar, but not identical \citep{Greenland2000}. Usually, the assumption of exchangebility indicates the schedule-specific parameters come from a common distribution with an overall mean. For the $\text{ED}_{50}$ parameter, we assume the re-scaled and log transformed $\text{ED}_{50}$ parameter estimates (using the corresponding frequency of each schedule) are exchangeable.

In this manuscript, we propose a Bayesian hierarchical model which utilizes shrinkage estimation for certain parameters of the dose-response model in order to dynamically borrow strength across schedules in a phase II trial. Another contribution is the introduction of a publicly available R package, \texttt{ModStan}. In Section~\ref{sec:app}, two phase II trials with multiple schedules for the treatment of atopic dermatitis are described. We introduce the proposed method to analyze phase II trials with multiple schedules in Section~\ref{sec:method}. We also describe partial pooling with assuming schedule-specific fixed-effects for certain parameters, discuss the choice of priors and implementation of the proposed method. We evaluated the long-run properties of different methods in a simulation study in Section~\ref{simulations}. One of the illustrative applications is revisited to display the proposed method and compare it to the alternatives in Section~\ref{Dupi}. We close with some conclusions and outlook.

%%%%%%%%%%%%%%%%%%%%%%%%%%%%%%%%%%%%%%%%%%%%%%%%%%%%%%
\section{Illustrative applications}\label{sec:app}
%%%%%%%%%%%%%%%%%%%%%%%%%%%%%%%%%%%%%%%%%%%%%%%%%%%%%%
Atopic dermatitis, the most common form of eczema, is a chronic inflammatory 
disease that is characterized by skin rash and itching \citep{atopic}. Recently, there is an increasing number of clinical trials investigating novel systemic agents for the treatment of atopic dermatitis \citep{alexander2019}.
We consider phase II trials of two human monoclonal antibodies, dupilumab and MOR106, for the treatment
of atopic dermatitis. Designs of two trials are listed in Table~\ref{tab:Data}. For both trials, patients were randomized
into six arms including a placebo arm. We consider these two trials, since they were both designed to investigate
multiple schedules. The dupilumab trial contains three schedules 
(weekly, biweekly, and monthly), whereas the MOR106 trial contains two (biweekly and monthly). The placebo doses were administered with the highest frequencies including a weekly and a biweekly schedule for dupilumab and MOR106, respectively. The primary endpoint of both trials is the percentage change from baseline in Eczema Area and Severity Index (EASI) score at Day 85. 
The EASI scoring system is used to grade the
severity of the signs of eczema. EASI scores take values between 0 and
72 and higher EASI score means higher severity. Dupilumab and MOR106 trials are used to motivate our simulation studies in Section~\ref{simulations}. The dupilumab trial was completed in September 2014. Multiple comparisons procedure was used as the primary statistical analysis in the dupilumab trial \citep{dupilumab}. For the purpose of illustration, we will analyze the dupilumab trial using different modeling approaches in Section~\ref{Dupi}. In October 2019, the MOR106 trial was terminated due to lack of efficacy in the interim analysis \citep{MOR106}.

\begin{table}[htb]
\centering
\caption{Designs of two phase II trials in atopic dermatitis (Dupilumab and MOR106) involving different schedules. Clinicaltrial.gov identifiers are displayed for two trials.}
\label{tab:Data}
\begin{tabular}{lcccccc}
  \toprule
   &\multicolumn{3}{c}{\textbf{Dupilumab}: NCT01859988}     & \multicolumn{3}{c}{\textbf{MOR106}: NCT03568071}  \\
   \cmidrule(rl){2-4} \cmidrule(rl){5-7}
  	            Arm                 & Schedule & Dose & Planned  & Schedule & Dose  & Planned  \\
  	                                   &                       &   (mg/m$^2$)  & sample size &                   & (mg/kg)      & sample size \\ \midrule
                1 & Weekly      & 0       & 40 & Biweekly  & 0 & 45\\
                2 & Weekly      & 300   & 40 & Biweekly & 1 & 45\\
                3 & Biweekly & 200   & 40  & Biweekly & 3 & 45\\
                4 & Biweekly & 300   & 40  & Biweekly & 10 & 45\\
                5 & Monthly    & 100  & 40  & Monthly & 1 & 30\\
                6 & Monthly    & 300  & 40  & Monthly & 3 & 30\\
   \bottomrule
\end{tabular}
\end{table}

%%%%%%%%%%%%%%%%%%%%%%%%%%%%%%%%%%%%%%%%%%%%%%%%%%%%%%
\section{Statistical methods}\label{sec:method}
%%%%%%%%%%%%%%%%%%%%%%%%%%%%%%%%%%%%%%%%%%%%%%%%%%%%%%
Assume that a response \(y_{ijk}\) (an efficacy or a safety outcome) is observed for schedule $i$, dose $j$ and patient $k$. Following \cite{Feller2017}, we assume a normal likelihood for a continuous outcome:
\begin{align}
y_{ijk}\sim  \mathcal{N}(f(d_{j}^{(i)}, \bm{\theta}), \sigma_{i}^2) \label{model:normal}
\end{align}
where \(\bm{\theta}\) refers to the model parameters and \(\sigma_{i}\) to
the error standard deviation. The $f(d_{j}^{(i)}, \bm{\theta}) $ represents the functional form of the dose-response relationship for schedule $i$. Other outcome types, for instance dichotomous or count, can be modeled by specifying appropriate likelihood (e. g. Binomial or Poisson) and the link function (e. g. logit or log transformation).

There are a number of candidate models for the functional form including the popular Emax model \citep{Thomas2014}, that is
\begin{align}
f(d_{j}^{(i)}, \bm{\theta}) = \text{E}_{0}^{(i)} + \text{E}_{\text{max}}^{(i)}\,\frac{d_{j}^{(i)}}{\text{ED}_{50}^{(i)} + d_{j}^{(i)}} \label{model:Emax}
\end{align}
where $\text{E}_{0}^{(i)}$ is the placebo response and $\text{E}_{\text{max}}^{(i)}$ is the maximum effect attributable to the drug. The $\text{ED}_{50}^{(i)}$ parameter represents the dose at which half of the maximum effect is reached. In the manuscript, we exclusively use the Emax model, see \cite{Bretz2005} for different candidate models.

As explained in the introduction, one way of modeling the dose-response curves is to treat all model parameters as schedule-specific fixed-effects. However, such an analysis is not the most efficient, when certain aspects of the dose-response curves in different schedules are similar. Alternatively, one can consider a complete pooled analysis in which all model parameters from different schedules are assumed to be the same. This approach is also problematic, since it ignores the potential heterogeneity between dose-response curves of different schedules. A more reasonable approach is the partial pooling \citep{Feller2017}, which strikes a balance between efficiency and robustness. It is often reasonable to assume that placebo effect \(\text{E}_{0}^{(i)}\) is the
same for different schedules, that is,
$\text{E}_{0}^{(1)} = \text{E}_{0}^{(2)} = \dots$. This is especially the case, when there is only one placebo arm investigated in the trial as in the illustrative trials described in Section~\ref{sec:app}. In some situations, it might also make sense to assume that the maximum efficacy for high doses is same, $\text{E}_{\text{max}}^{(1)} = \text{E}_{\text{max}}^{(2)} = \dots$. However, it might not be reasonable to assume the dose providing half of the maximum efficacy is the same for different schedules, that is $\text{ED}_{50}^{(1)} \neq \text{ED}_{50}^{(2)} \neq \dots$. \cite{Feller2017} suggested to treat the unshared parameters, for example $\text{E}_{\text{max}}^{(i)}$ and/or $\text{ED}_{50}^{(i)}$, as schedule-specific fixed-effects in the partial pooling approach. 

%%%%%%%%%%%%%%%%%%%%%%%%%%%%%%%%%%%%%%%%%%%%%%%%%%%%%%%%%%%%%%%%%%%%%%%%%%
\subsection{Proposed method: Partial pooling with random-effects}\label{sec:shrinkage}
%%%%%%%%%%%%%%%%%%%%%%%%%%%%%%%%%%%%%%%%%%%%%%%%%%%%%%%%%%%%%%%%%%%%%%%%%%
Rather than using schedule-specific fixed-effects, we propose a Bayesian hierarchical model with random-effects to model the between-schedule heterogeneity with regard to certain parameters in the partial pooling approach. In other words, we suggest partial pooling with assuming schedule-specific random-effects for certain parameters of the dose-response model.
To be concrete, assume that we want to obtain schedule-specific random-effects for $\text{ED}_{50}^{(i)}$. Firstly, we need to re-scale $\text{ED}_{50}^{(i)}$ parameters. For this reason, we specify a \emph{reference schedule} (\(i_{\text{ref}}\)). The re-scaled parameters are given by $\text{ED}_{50}^{*(i)} = \text{ED}_{50}^{(i)} \, (f^{(i)}/f^{(i_{\text{ref}})})$ where $f^{(i_{\text{ref}})}$ and $f^{(i)}$ are the frequency of administration of the reference schedule $i_{\text{ref}}$ and the schedule $i$, respectively. The $\text{ED}_{50}^{(i)}$ is modeled on the log-scale, since it is necessarily positive as a dose. We assume that the re-scaled schedule-specific $\text{ED}_{50}^{*(i)}$ estimates are exchangeable
\begin{align}
 \text{log}(\text{ED}_{50}^{*(i)}) \sim \mathcal{N}(\mu_{\text{ED}_{50}}, \tau_{\text{ED}_{50}}^2) \label{eq:shrinkage1}
\end{align}
where \(\mu_{\text{ED}_{50}}\) is the overall mean and $\tau_{\text{ED}_{50}}$ is the between-schedule heterogeneity in $\text{log}(\text{ED}_{50}^{*(i)})$. Our main interest is in the schedule-specific estimates, $\text{ED}_{50}^{(i)}$. If the heterogeneity $\tau_{\text{ED}_{50}}$ is zero, then the model reduces to a model assuming shared $\text{ED}_{50}^{*(i)}$ parameters. Note that the results are invariant to the choice of the reference schedule. Furthermore, similar to the \(\text{ED}_{50}\) parameter, shrinkage estimates of \(\text{E}_{\text{max}}\) parameter can be obtained. There is no need to use the re-scaling or the log transformation for the \(\text{E}_{\text{max}}\) parameter. Treating $\text{E}_{\text{max}}$ and/or $\text{ED}_{50}$ parameters differently, assuming either one or both of them shared between schedules or assuming schedule-specific random-effects, results in a variety of alternative models.

Complete pooling and partial pooling approaches can be fitted using 
likelihood estimation. For example, \cite{Mollenhof2019} demonstrated the likelihood implementation of the partial pooling with assuming schedule-specific fixed-effects for $\text{ED}_{50}^{(i)}$ using constrained nonlinear optimization via \texttt{alabama}  \citep{alabama} R package. Alternatively, Bayesian approaches can be used, which we consider in this paper.

%%%%%%%%%%%%%%%%%%%%%%%%%%%%%%%%%%%%%%%%%%%%%%%%%%%%%%%%%%
\subsection{Prior distributions}\label{sec:priors}
%%%%%%%%%%%%%%%%%%%%%%%%%%%%%%%%%%%%%%%%%%%%%%%%%%%%%%%%%%
For the Bayesian implementation, we need to specify prior distributions for the model parameters $\text{E}_{0}$, $\text{E}_{\text{max}}$ $\mu_{\text{ED}_{50}}$,  \(\tau_{\text{ED}_{50}}\) and $\sigma$ for the partial pooling assuming schedule-specific random-effects for $\text{ED}_{50}^{(i)}$. We use vague (non-informative) priors, \(\mathcal{N}(0, 100^2)\), for the parameters $\text{E}_{0}$ and $\text{E}_{\text{max}}$, and a half-normal prior with scale 100 for $\sigma$, \(\mathcal{HN}(100)\). The parameters $\mu_{\text{ED}_{50}}$ and \(\tau_{\text{ED}_{50}}\) need special attention, since the priors of both parameters have strong influence on the posterior estimates. The difficulty of the estimation of the $\tau_{\text{ED}_{50}}$ stems from the small number of the schedules. For example, in our two illustrative trials, there are only two and three schedules available. This is similar to the meta-analysis of few studies, in which the estimation of the between-trial heterogeneity has gained considerable attention in the literature \citep{gelman2006}. \cite{friede2017} suggested the use of \emph{weakly informative priors} (WIP) for the heterogeneity parameter in the case of meta-analysis of few studies, specifically half-normal priors with the scale of 0.5 or 1, when relative measures such as odds ratios, relative risks or hazard ratios (on the logarithmic scale) are used to describe the effect. Inspired by these, we can also construct a WIP for the $\tau_{\text{ED}_{50}}$ to represent plausible range of $\text{ED}_{50}^{*(i)}$ values \citep{spiegelhalter2004}. The 95\% of values of $\text{log}(\text{ED}_{50}^{*(i)})$ will lie in the interval $\mu_{\text{ED}_{50}} \pm 1.96\cdot \tau_{\text{ED}_{50}}$, hence the 97.5\% and 2.5\% values of $\text{log}(\text{ED}_{50}^{*(i)})$ are $2 \cdot 1.96 \cdot \tau_{\text{ED}_{50}}$ apart. Accordingly, the ratio of the 97.5\% to the 2.5\% point of the distribution of $\text{ED}_{50}^{*(i)}$ values is exp($3.92 \cdot \tau_{\text{ED}_{50}}$). Table~\ref{tab:heterogeneity} lists the ``range'' of $\text{ED}_{50}^{*(i)}$ values based on different $\tau_{\text{ED}_{50}}$. In order to cover typical $\tau_{\text{ED}_{50}}$ values conservatively, we will use half-normal priors with scale 1, i.e. $\mathcal{HN}(1)$. When we are interested in the shrinkage estimates of $\text{E}_{\text{max}}$, the construction of the WIP for $\tau_{\text{Emax}}$ is slightly different. This is because $\text{E}_{\text{max}}$ is computed on the original scale, not on the logarithmic scale. Here, the difference (instead of the ratio) between the 97.5\% and the 2.5\% point of the distribution of $\text{E}_{\text{max}}$ values is $3.92 \cdot \tau_{\text{Emax}}$. To cover plausible $\tau_{\text{Emax}}$ values, we will use half-normal priors with the scale 10, $\mathcal{HN}(10)$.

\begin{table}[htb]
\centering
\caption{Between-schedule heterogeneity $\tau_{\text{ED}_{50}}$ in $\text{log}(\text{ED}_{50}^{*(i)})$: $\tau_{\text{ED}_{50}}$ referring small to very large heterogeneity. The ``range'', exp($3.92 \cdot \tau_{\text{ED}_{50}}$), refers to the ratio of the 97.5\% to the 2.5\% point of the distribution of $\text{ED}_{50}^{*(i)}$.}
\label{tab:heterogeneity}
\begin{tabular}{lccccc}
  \toprule
  	            $\tau_{\text{ED}_{50}}$               &    ``range'' of  $\text{ED}_{50}^{*(i)}$   \\        
  	            \midrule
                0.125 (small)       &  1.63         \\ 
                0.25 (moderate)  &  2.66         \\ 
                0.5 (substantial)  &  7.10         \\ 
                1 (large)             &  50.40         \\ 
                2 (very large)      &  2540.20         \\ 
   \bottomrule
\end{tabular}
\end{table}

The parameter \(\text{ED}_{50}\) is different
from \(\text{E}_{0}\) and \(\text{E}_{\text{max}}\) in the sense that it is the only parameter that enters the model non-linearly. In the frequentist framework, it is a common practice to impose bounds (lower and upper bounds) on the space for \(\text{ED}_{50}\), since the maximum likelihood estimator (MLE) often does not converge \citep{Bornkamp2014}. However, the estimate will often exactly equal to the specified upper bound, which is unacceptable. In a Bayesian framework, simple prior choice for the \(\text{ED}_{50}\) are uniform distributions with finite bounds. However, uniform prior distributions on \(\text{ED}_{50}\) are problematic, since they strongly depend on the
parametrization: One may end up with completely different implied prior
distributions for the dose-response curve. A better prior for \(\text{ED}_{50}\) is the Jeffreys prior, which is invariant
to parametrization. It is defined as
$p(\bm{\theta}) \propto \sqrt{|I(\bm{d},\bm{w},\bm{\theta})|}$ where
\(\sqrt{|I(\bm{d},\bm{w},\bm{\theta})|}\) is the Fisher information, and
\(\bm{w}\) is the vector of proportion of patients allocated at dose
\(\bm{d}\). Hence, Jeffreys prior depends on the observed design
\((\bm{x}, \bm{w})\). One cannot state the Jeffreys prior before
data collection, which is crucial in many applications, e.g.~in the
presence of missing data or two stage designs.

\cite{Bornkamp2012} introduced the \emph{functional uniform prior}
which is a modified version of the Jeffreys prior. Functional
uniform priors are uniformly distributed on the potential different
shapes of the underlying nonlinear model function. These priors are also
invariant with respect to parametrization of the model function and
typically result in rather non-uniform prior distributions on the
parameter scale. Instead of the actual observed design, functional
uniform priors are calculated using a grid of doses as
\(\bm{x}\) and equal weights for \(\bm{w}\). More specifically, say, the
gradient function of the Emax model is given by
\(J_{x}(\bm{\theta}) = (1, x/(x+\text{ED}_{50}), -x/\text{E}_{\text{max}}/(x+\text{ED}_{50})^{2})\).
Let \(\bm{x}\) be a grid of doses and \(F(\bm{\theta})\) be the matrix
with \(J_{x}(\bm{\theta})\), \(x\) in the rows. Then, the functional
uniform prior is proportional to $\sqrt{|Z^{*}(\bm{\theta})|}$ where
\(Z^{*}(\bm{\theta}) = F^{T}(\bm{\theta})\,F(\bm{\theta})\) (see
\cite{Bornkamp2014} for more detailed explanations). An approximation of the functional uniform prior for \(\text{ED}_{50}\) is given as the log-normal distribution with mean -2.5 and standard deviation 1.8, when the \(\text{ED}_{50}\) is re-scaled with the maximum available dose $D$, that is $\text{ED}_{50} / D$ \citep{Bornkamp2014}. For the simulations and the application, we used the approximation of the functional uniform prior, since it is computationally cheaper. In all models, we use the bounds [0, $1.5 \cdot D$] for the space of \(\text{ED}_{50}\) (or $\mu_{\text{ED}_{50}}$) parameter.

%%%%%%%%%%%%%%%%%%%%%%%%%%%%%%%%%%%%%%%%%%%%%%%%%%%%%%
\subsection{Implementation of the proposed method}\label{sec:software}
%%%%%%%%%%%%%%%%%%%%%%%%%%%%%%%%%%%%%%%%%%%%%%%%%%%%%%
In a Bayesian framework, we fitted the described statistical models using the probabilistic programming language \textbf{Stan} which employs a modern Markov chain Monte Carlo (MCMC) algorithm, namely, Hamiltonian Monte Carlo with the No-U-Turn Sampler \citep{Stan}. The parametrization used for the statistical model influences the MCMC performance. A centered parametrization such as Equation~\eqref{eq:shrinkage1} may cause some computational difficulties such as difficulty in convergence in the presence of data sparsity such as meta-analysis of few studies \citep{betancourt2015hamiltonian} or dose-response modeling of phase II trials with few schedules. An alternative parametrization, that is a non-centered parametrization, overcomes these computational difficulties. To be more precise, by the reparametrization of the location and scale parameters, Equation~\eqref{eq:shrinkage1} becomes $ \text{log}(\text{ED}_{50}^{*(i)}) = \mu_{\text{ED}_{50}} + u_{i} \cdot \tau_{\text{ED}_{50}}$ where $u_{i} \sim \mathcal(0, 1)$ \citep{Guenhan2020Meta}. The \textbf{Stan} code defining the partial pooling with schedule-specific random-effects for $\text{ED}_{50}^{(i)}$ is shown in Listing~\ref{fig:Stan_PP}.

To facilitate the implementation of our proposed method for the practitioners, we have developed an R package, \texttt{ModStan} (\url{https://github.com/gunhanb/ModStan}). \texttt{ModStan} is a purpose-build package defined on top of the \texttt{rstan}, the R interface for \textbf{Stan}. We show how to install and use \texttt{ModStan} in Appendix~\ref{app1}. 

\begin{listing}
\begin{Verbatim}[numbers=left,frame=single,fontfamily=courier,fontsize=\footnotesize]

data {
  int<lower=1> N_obs;                           // num of observations 
  int<lower=1> N_schedule;                      // num of schedules
  int<lower=1> N_pred;                          // num of predicted doses
  real resp[N_obs];                             // responses 
  real<lower=0> dose[N_obs];                    // doses   
  int schedule[N_obs];                          // schedule indicator        
  real<lower=0> freq[N_obs];                    // frequency of administration (hrs)      
}
parameters {
  real E0;                                      // placebo effect (shared)
  real Emax;                                    // Emax parameter (shared)
  real log_ED50_raw[N_schedule];                // re-scaled log(ED50) parameters                   
  real<lower=0> sigma;                          // standard deviation for errors
  real<lower=0, upper=1.5> mu_ED50_raw;         // mean of log(ED50) random-effects
  real<lower=0> tau_ED50;                       // between-schedule heterogeneity
}
transformed parameters{
  real mu_ED50;
  real log_ED50[N_schedule];          
  real<lower=0> ED50[N_schedule];        
  vector[N_obs] resp_hat;
  
  mu_ED50 = log(mu_ED50_raw * max(dose));
  for(i in 1:N_schedule)
    log_ED50[i] = mu_ED50 + log_ED50_raw[i] * tau_ED50;
  // Taking exponentials and rescaling ED50 parameters
  for(i in 1:N_schedule)
    ED50[i] = exp(log_ED50[i]) * (freq[i]/ freq_ref);
  
  // Dose-response: Emax model
  for(i in 1:N_obs)  
    resp_hat[i] = E0 + (Emax * dose[i]) / (ED50[schedule[i]] + dose[i]);
}
model {
  // random-effects
  log_ED50_raw ~ normal(0, 1);  // implies log(ED50) ~ normal(mu_ED50, tau_ED50)
  // likelihood
  resp ~ normal(resp_hat, sigma);
  // prior distributions
  sigma  ~ normal(0, 100); 
  E0 ~ normal(0, 100); 
  Emax ~ normal(0, 100);
  // approximation to the functional uniform prior
  mu_ED50_raw  ~ lognormal(-2.5, 1.8); 
  tau_ED50  ~ normal(0, 1);  
}
\end{Verbatim}
\caption{\textbf{Stan} code defining the partial pooling with schedule-specific random-effects for $\text{ED}_{50}$ parameter. The parameters \(\text{E}_{0}\), \(\text{E}_{\text{max}}\) and $\sigma$ are assumed to be shared between schedules.}\label{fig:Stan_PP}
\end{listing}

%%%%%%%%%%%%%%%%%%%%%%%%%%%%%%%%%%%%%%%%%%%%%%%%%%%%%%
\section{Simulation study}\label{simulations}
%%%%%%%%%%%%%%%%%%%%%%%%%%%%%%%%%%%%%%%%%%%%%%%%%%%%%%
In order to evaluate the long-run properties of the proposed method and compare it with some alternative methods, a simulation study was conducted.

%%%%%%%%%%%%%%%%%%%%%%%%%%%%%%%%%%%%%%%%%
\subsection{Simulation settings and implementation}\label{settings}
\begin{figure}[htb]
\centering
\includegraphics[scale=0.5]{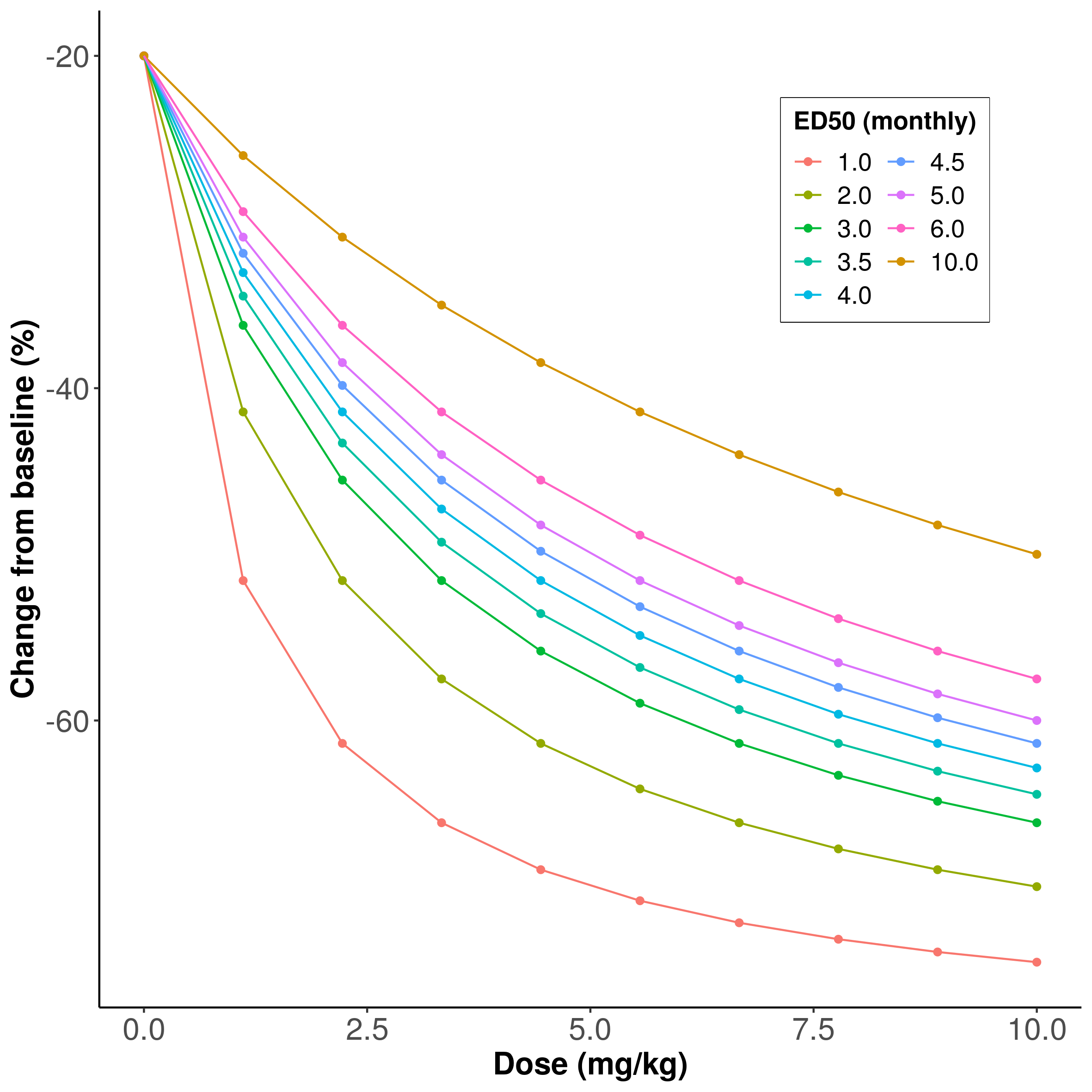}
\caption{Dose-response curves for the monthly schedule investigated in the simulation study. Different curves are generated by varying $\text{ED}_{50}^{\text{monthly}}$ parameter value.\label{fig:scenarios}}

\end{figure}

%%%%%%%%%%%%%%%%%%%%%%%%%%%%%%%%%%%%%%%%%
The scenarios considered are motivated by the dupilumab and MOR106 trials described in Section~\ref{sec:app}. Each generated trial consists of seven arms: one placebo arm and 1, 3, and 10 mg/kg for both biweekly and monthly schedules. The primary outcome is the percentage change from baseline in EASI score. Hence, the datasets are generated under the assumption of normally distributed outcomes, specifically Equation~\eqref{model:normal}. The underlying dose-response function is assumed to be an Emax model, that is Equation~\eqref{model:Emax}. True values for $\text{E}_{0}^{(i)}$, $\text{E}_{\text{max}}^{(i)}$ and $\sigma_{i}$ are taken as $-20\%$, $-60\%$, and $35\%$ for both schedules, respectively. Furthermore, $\text{ED}_{50}^{\text{biweekly}}$ is assumed to be 2 mg/kg. A total of 27 scenarios are obtained by varying the $\text{ED}_{50}^{\text{monthly}}$ ($\text{ED}_{50}^{\text{monthly}} \in \{1, 2, 3, 3.5, 4, 4.5, 5, 6, 10$ (mg/kg)$\}$) and sample sizes of each arm ($N \in \{30, 45, 60\}$). $\text{ED}_{50}^{\text{monthly}}$ values are chosen to investigate the influence of the difference between true values of $\text{ED}_{50}^{\text{biweekly}}$ and $\text{ED}_{50}^{\text{monthly}}$ on the performance. Figure~\ref{fig:scenarios} displays different dose-response curves for the monthly schedule investigated in the simulations. When $\text{ED}_{50}^{\text{monthly}}$ corresponds to 4 mg/kg, there is no heterogeneity in $\text{ED}_{50}$ parameters between schedules. This is because if we re-scale $\text{ED}_{50}^{\text{monthly}}$ to transform on the biweekly scale (simply dividing by two), we obtain 2 mg/kg, which is the true value of $\text{ED}_{50}^{\text{biweekly}}$. Accordingly, when the true value of $\text{ED}_{50}^{\text{monthly}}$ deviates from 4 mg/kg, the heterogeneity between schedules in $\text{ED}_{50}$ increases. The simulations were carried out with \mbox{$1\,000$} replications per scenario.

In the proposed method, we assume that $\text{E}_{0}^{(i)}$, $\text{E}_{\text{max}}^{(i)}$ and $\sigma_{i}$ are shared between schedules, while $\text{ED}_{50}^{(i)}$ are assumed to be schedule-specific random-effects. In other words, the proposed method corresponds to the partial pooling with assuming schedule-specific random-effects for $\text{ED}_{50}^{(i)}$ (``PP - RE''). As a comparator, we use the model in which $\text{ED}_{50}^{(i)}$ are assumed to be schedule-specific fixed-effects, while other parameters are shared (``PP - FE''). Both partial pooling approaches (PP - RE and PP - FE) are fitted via a Bayesian approach. We also consider the complete pooling method via a frequentist and a Bayesian approach (``CP (Frequentist)'' and ``CP (Bayesian)''). For the partial pooling with schedule-specific random-effects, we used the biweekly schedule as the reference schedule to re-scale the $\text{ED}_{50}^{(i)}$ parameters. To implement the complete pooling approaches, all doses should be re-scaled into the same schedule. For this purpose, we transform the doses from the monthly schedule into the biweekly schedule. Accordingly, the new set of doses becomes $\{0, 0.5, 1, 1.5, 3, 10$ (mg/kg)$\}$ for complete pooling approaches.

The complete pooling (Frequentist) is fitted using \texttt{fitMod} function from the \texttt{DoseFinding} \citep{DoseFinding} R package. All Bayesian methods are fitted using \textbf{Stan} and the prior distributions from Section~\ref{sec:priors} are used. Three MCMC chains were run in parallel for a total of \mbox{$4\,000$} iterations including \mbox{$2\,000$} iterations of burn-in. Convergence diagnostics are evaluated in some replications, these MCMC settings are chosen accordingly. The $\text{ED}_{50}$ parameter is assumed to be within the bounds of  [0.001, 1.5 $\cdot$ 10] to ensure identifiability for all methods.

%%%%%%%%%%%%%%%%%%%%%%%%%%%%%%%%%%%%%%%%%%%%%%%
\subsection{Simulation results}\label{settings}
%%%%%%%%%%%%%%%%%%%%%%%%%%%%%%%%%%%%%%%%%%%%%%%
For each simulation run, we calculated point estimates ($\hat{f}$) for the dose-response function ($f$) (the pointwise posterior median or the maximum likelihood estimate) at some pre-specified dose levels. For this purpose, ten dose levels are chosen between 0 and 10 mg/kg equidistantly, namely $\text{dose}_{l} \in \{0.00, 1.11, \ldots , 10.00\}$. Additionally, interval estimates (95\% confidence interval or 95\% equi-tailed credible intervals) are derived at each $\text{dose}_{l}$. These computations are done for the dose-response function of the biweekly schedule. The following three performance measures are calculated:
\begin{itemize}
\item MAE: Mean absolute error for the dose-response function,
  \(1/10 \, \Sigma_{\text{dose}_{l}=0}^{10}|f(\text{dose}_{l}) - \hat{f}(\text{dose}_{l})|\) at each $\text{dose}_{l}$.
\item Coverage probability: Mean coverage probability of the interval estimates 
  evaluated at each $\text{dose}_{l}$.
\item Mean length: Mean length of the interval estimates at each $\text{dose}_{l}$.
\end{itemize}

\begin{figure}[htb]
\centering
\includegraphics[scale=0.45]{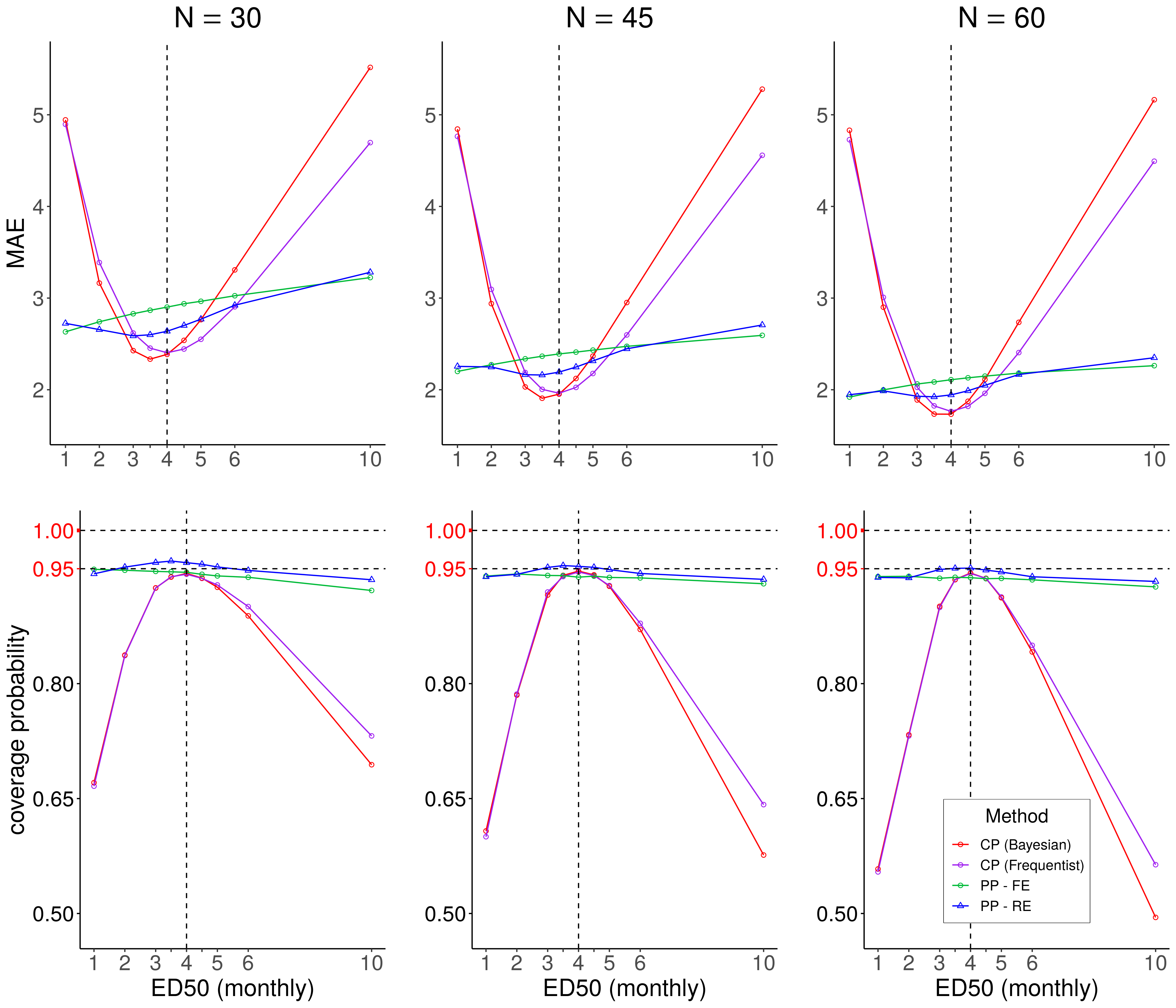}
\caption{Simulation results for different sample sizes $H$ per arm. The mean absolute error (MAE) and coverage probabilities for the dose-response curve obtained by four methods with different sample sizes. Four methods include complete pooling approaches using frequentist, CP (Frequentist), and Bayesian methods, CP (Bayesian), and partial pooling approaches using schedule-specific fixed-effects (PP - FE) and schedule-specific random-effects (PP - RE) for $\text{ED}_{50}^{(i)}$. The vertical dashed line indicates the scenario without heterogeneity in the re-scaled $\text{ED}_{50}^{(i)}$ between biweekly and monthly schedules. $\text{ED}_{50}^{\text{biweekly}}$ is assumed to be 2 mg/kg.\label{fig:res}}
\end{figure}

The lower MAE for the point estimates, the shorter interval estimates, and the coverage probability of 95\% for the interval estimates are desirable. The MAE obtained by the four methods is displayed in the first row of Figure~\ref{fig:res}. Different columns of Figure~\ref{fig:res} correspond to different sample sizes $N$ which are investigated in the simulations. Across different sample sizes, the relative performances of the four methods remain similar. The scenario of $\text{ED}_{50}^{\text{monthly}} = 4$ corresponds to the scenario without heterogeneity in the re-scaled $\text{ED}_{50}^{(i)}$ between biweekly and monthly schedules, which is shown by a vertical dashed line. The heterogeneity increases, when $\text{ED}_{50}^{\text{monthly}}$ deviates from 4. Both complete pooling approaches display better performance than both partial pooling approaches in terms of the MAE, when the $\text{ED}_{50}^{\text{monthly}}$ is 4. However, the partial pooling approaches result in more robust performance across $\text{ED}_{50}^{\text{monthly}}$ values in comparison to the pooling approaches. The partial pooling with random-effects uses the prior $\mathcal{HN}(1)$ for the heterogeneity parameter $\tau_{\text{ED}_{50}}$. If we increase the value of the prior standard deviation (that is 1), then the performance of the partial pooling with random-effects will get closer to the partial pooling with fixed-effects. Similarly, if we assume that $\tau_{\text{ED}_{50}}$ equals to zero, the partial pooling with random-effects reduces to, effectively, the complete pooling (Bayesian). The partial pooling with random-effects yields better performance than the partial pooling with fixed-effects in terms of the MAE across different $\text{ED}_{50}^{\text{monthly}}$ values and sample sizes except the most extreme scenarios, namely $\text{ED}_{50}^{\text{monthly}} = 1$ or 10. Note that the main difference between the complete pooling (Bayesian) and complete pooling (Frequentist) is that in the former, functional uniform priors used for $\text{ED}_{50}^{(i)}$ parameters. The small discrepancy between the MAE obtained by the complete pooling (Bayesian) and the complete pooling (Frequentist) can be explained by this difference. Furthermore, when the sample sizes increase, the MAE decreases in the four methods as expected.

\begin{figure}[htb]
\centering
\includegraphics[scale=0.11]{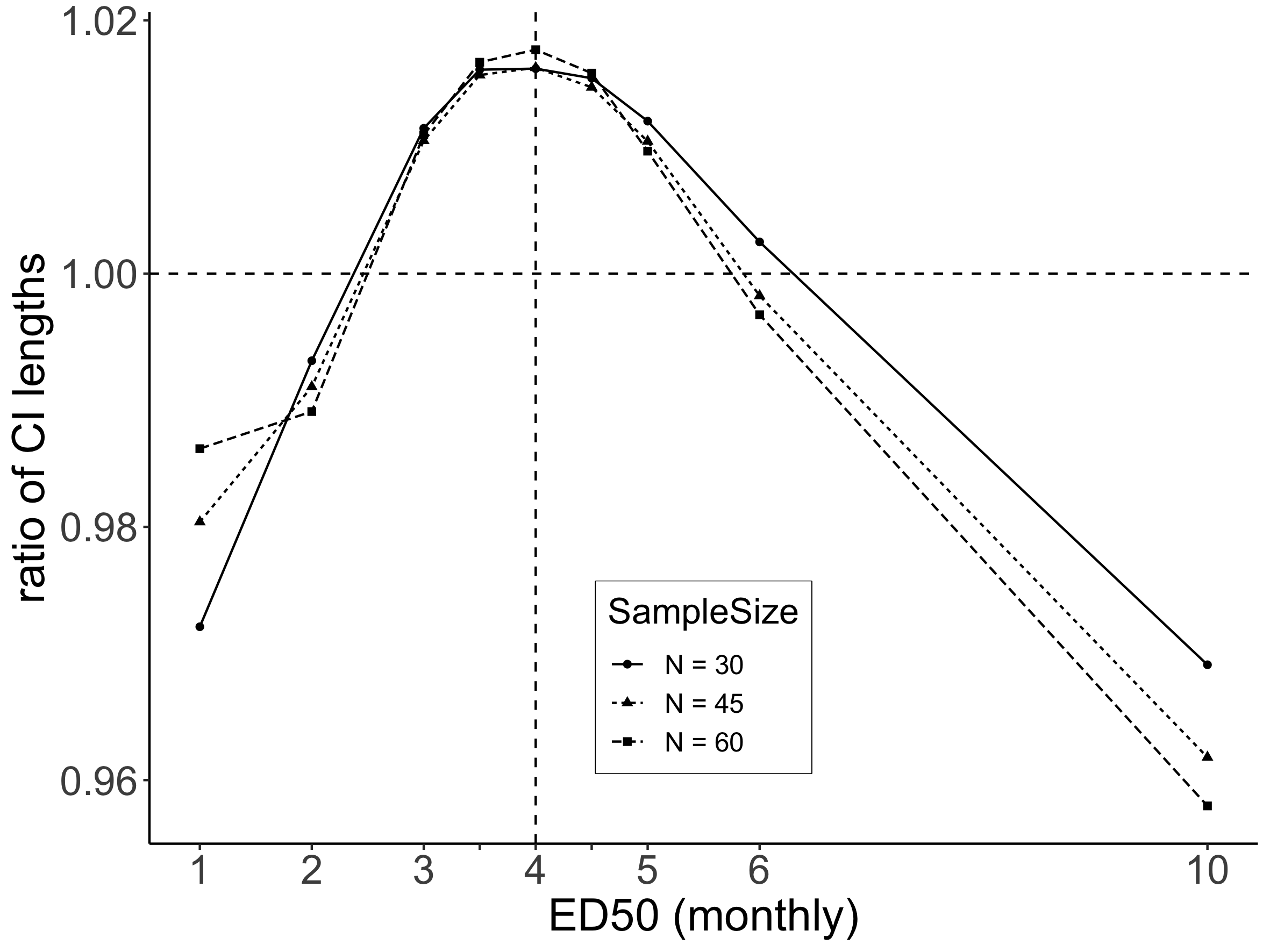}
\caption{Simulation results for different sample sizes. Ratios of lengths of credible intervals for the dose-response curves obtained by the partial pooling with random-effects and the partial pooling with fixed-effects with different sample sizes. The denominator of the ratio is the length of credible interval obtained by the partial pooling with random-effects.  The vertical dashed line indicates the scenario without heterogeneity in the re-scaled $\text{ED}_{50}^{(i)}$ between biweekly and monthly schedules. $\text{ED}_{50}^{\text{biweekly}}$ is assumed to be 2 mg/kg.\label{fig:CI}}
\end{figure}

Figure~\ref{fig:res} also shows coverage probabilities of the interval estimates obtained by the four methods. The complete pooling approaches result in a concave shape and display unacceptably low coverage when $\text{ED}_{50}^{\text{monthly}}$ deviates from 4. This undesirable performance of the complete pooling approaches is more pronounced, when the sample size increases. As in the MAE, both partial pooling approaches show more robust performance in terms of the coverage probabilities in comparison to the complete pooling approaches. The partial pooling with random-effects yields superior performance in terms of the coverage probability compared to the partial pooling with fixed-effects across different $\text{ED}_{50}^{\text{monthly}}$ values and sample sizes except when $\text{ED}_{50}^{\text{monthly}} = 1$. Figure~\ref{fig:CI} illustrates the ratios of lengths of credible intervals for the dose-response functions obtained by the partial pooling approaches. The denominator of the ratio is the length of the credible interval obtained by the partial pooling with random-effects. The partial pooling with random-effects results in slightly shorter credible intervals, while it produces slightly higher coverage probability compared to the partial pooling with fixed-effects in most of the scenarios.

To examine the influence of the potential heterogeneity in $\text{E}_{\text{max}}^{(i)}$ between schedules, we conducted additional simulations. True values for $\text{E}_{0}^{(i)}$ and $\sigma_{i}$ are taken as -20\% and 35\% for both schedules, respectively. The $\text{ED}_{50}^{\text{biweekly}}$ and $\text{ED}_{50}^{\text{monthly}}$ are assumed to be 2 and 4 mg/kg, respectively. This corresponds to assuming no heterogeneity in $\text{ED}_{50}^{(i)}$ parameters between schedules, since we focused on $\text{E}_{\text{max}}^{(i)}$ in these simulations. The $\text{E}_{\text{max}}^{\text{biweekly}}$ is assumed to be -60\%. Sample size for each arm is 45. Three scenarios are generated by varying $\text{E}_{\text{max}}^{\text{monthly}}$ values ($\text{E}_{\text{max}}^{\text{monthly}} \in \{-70\%, -60\%, -50\%\}$). Notice that when $\text{E}_{\text{max}}^{\text{monthly}} = -60\%$, there is no heterogeneity in $\text{E}_{\text{max}}^{(i)}$ values. In the partial pooling with fixed-effects, both $\text{E}_{\text{max}}^{(i)}$ and $\text{ED}_{50}^{(i)}$ parameters are treated as schedule-specific fixed-effects. In the partial pooling with random-effects, both parameters $\text{E}_{\text{max}}^{(i)}$ and $\text{ED}_{50}^{(i)}$ are assumed to be schedule-specific random-effects. The simulation results are listed in Table~\ref{tab:EmaxVary}. In the scenario of $\text{E}_{\text{max}}^{\text{monthly}} = -60\%$, the complete pooling approaches result in lower MAE in comparison to the partial pooling approaches, while reaching the coverage probability of 95\% for the confidence intervals. However, in other scenarios, complete pooling approaches yield worse performance in terms of the MAE and coverage probabilities compared to the partial pooling approaches. The partial pooling with random-effects results in smaller MAE and the shorter credible intervals compared to the partial pooling with fixed-effects in all three scenarios. 

When we take into account all simulation results, the partial pooling approaches are more robust in terms of the MAE and the coverage probabilities across scenarios compared to the complete pooling approaches. The partial pooling with random-effects yields better performance than the partial pooling with fixed-effects in terms of the MAE and the mean length of the credible intervals with the exception of highly heterogeneous scenarios.

\begin{table}[htb]
\centering
\caption{Simulation results for varying $\text{E}_{\text{max}}^{\text{monthly}}$ scenarios. The mean absolute error (MAE) for the dose-response function, coverage probabilities and mean length of the interval estimates for the dose-response function obtained by the four methods. Four methods include complete pooling approaches using frequentist, CP (Frequentist), and Bayesian methods, CP (Bayesian), and partial pooling approaches using schedule-specific fixed-effects (PP - FE) and schedule-specific random-effects (PP - RE) for $\text{ED}_{50}^{(i)}$. }
\label{tab:EmaxVary}
\begin{tabular}{llll}
  \toprule
            &   \multicolumn{3}{c}{$\text{E}_{\text{max}}^{\text{monthly}}$} \\ 
  \midrule
           & -60\% & -70\% & -50\% \\
  \midrule  
            \multicolumn{4}{c}{Mean absolute error} \vspace{0.2cm}  \\
  CP (Frequentist)         & 1.63 & 2.44 & 2.51   \\
  CP (Bayesian)                  & 1.62 & 2.23 & 2.87   \\
  PP - FE                      & 2.04 & 2.03 & 2.03   \\
  PP - RE                      & 1.88 & 1.90 & 2.02   \\
  
            \multicolumn{4}{c}{Coverage probability} \vspace{0.2cm}    \\
  CP (Frequentist)         & 0.95 & 0.87 & 0.84   \\
  CP (Bayesian)                  & 0.95 & 0.87 & 0.84   \\
  PP - FE                      & 0.96 & 0.96 & 0.96   \\
  PP - RE                      & 0.96 & 0.96 & 0.95   \\
  
            \multicolumn{4}{c}{Mean length} \vspace{0.2cm}        \\
  CP (Frequentist)         & 5.78 & 5.77 & 5.82  \\
  CP (Bayesian)                 & 5.53 & 5.48 & 5.47   \\
  PP - FE                      & 6.96 & 6.96 & 6.95   \\
  PP - RE                     & 6.59  & 6.64 & 6.71   \\
   \bottomrule
\end{tabular}
\end{table}

%%%%%%%%%%%%%%%%%%%%%%%%%%%%%%%%%%%%%%%%%%%%%%%%%%%%%%
\section{Revisiting the Dupilumab trial}\label{Dupi}
%%%%%%%%%%%%%%%%%%%%%%%%%%%%%%%%%%%%%%%%%%%%%%%%%%%%%%
\begin{table}[htb]
\centering
\caption{The dupilumab trial: Sample sizes, least square (LS) means, and standard errors for each arm in the trial.}
\label{tab:Dupilumab}
\begin{tabular}{lccccc}
  \toprule
  	            Arm & Schedule & Dose (mg/$\text{m}^2$) & Sample size &  LS mean & Standard error         \\ \midrule
               1 &  Weekly  & 0      & 61   & -18.1  & 5.2\\ 
               2 & Weekly   & 300  & 63   & -73.7  & 5.2\\ 
               3 & Biweekly & 200  & 61   & -65.4  & 5.2\\ 
               4 & Biweekly & 300  & 64   & -68.2  & 5.1\\ 
               5 & Monthly & 100  & 65    &  -44.8 & 5.0\\ 
               6 & Monthly & 300  & 65    &  -63.5 & 4.9\\ 
   \bottomrule
\end{tabular}
\end{table}

We return to the dupilumab trial which was described in Section~\ref{sec:app}. The least square means and standard errors for different arms of the trial are listed in Table~\ref{tab:Dupilumab} as reported in \cite{dupilumab}. In total, 379 patients completed the trial. We analyzed the dataset assuming normal distribution for least square means with the given standard errors. Note that this is different than assuming normality for the observations as described in Equation~\eqref{model:normal} as reported in the reference paper for convenience \citep{dupilumab}. This will show that the proposed method also works with weaker assumption, as we only use an arm-level data instead of an observation-level data. Five different models were fitted in a Bayesian framework. We compare them via the approximate leave-one-out cross-validation information criteria (LOO-IC) \citep{Vehtari2017}. Note that LOO-IC has the same purpose as the Akaike Information Criteria (AIC) used in the frequentist framework and similar to the AIC, the lower value indicates the better model. All models assume an Emax model for the dose-response relationship. We use prior distributions described in Section~\ref{sec:priors}. The model descriptions are listed in Table~\ref{tab:LOOIC}. Model 1 corresponds to the complete pooling. In Models 2-5, the $\text{E}_{0}^{(i)}$ are assumed to be shared between schedules, while $\text{ED}_{50}^{(i)}$ and $\text{E}_{\text{max}}^{(i)}$ are treated differently in each model. Hence, Models 2-5 are partial pooling approaches. In Models 2 and 3, $\text{E}_{\text{max}}^{(i)}$ are assumed to be shared between schedules. Model 2 assumes schedule-specific fixed-effects for $\text{ED}_{50}^{(i)}$, while Model 3 uses schedule-specific random-effects for $\text{ED}_{50}^{(i)}$. Model 4 assumes schedule-specific fixed-effects both for $\text{ED}_{50}^{(i)}$ and $\text{E}_{\text{max}}^{(i)}$, whereas Model 5 uses schedule-specific random-effects both for $\text{ED}_{50}^{(i)}$ and $\text{E}_{\text{max}}^{(i)}$. For the complete pooling, the doses are transformed into the biweekly scale, thus the new set of doses are $\{0, 50, 150, 200, 300, 600\}$. For Models 3 and 5, we use the biweekly schedule as the reference schedule. 

Table~\ref{tab:LOOIC} displays the LOO-IC values for the five models. The complete pooling results in the best model in terms of the LOO-IC. The second and third best models are the partial pooling with schedule-specific random-effects for $\text{ED}_{50}^{(i)}$ and the partial pooling with schedule-specific fixed-effects for $\text{ED}_{50}^{(i)}$, respectively. Apparently, the model complexity is heavily penalized by LOO-IC for this dataset, hence LOO-IC results in lower values for the simpler models. One possible reason is the data sparsity, numbers of dose levels available for different schedules are 2, 3, and 3 (by including placebo arm for all schedules). Based on these results, hereafter, we focus on Models 1-3.

\begin{table}[htb]
\centering
\caption{Analyzing the dupilumab trial: The approximate leave-one-out information criterion (LOO-IC) obtained by five different models. In all models, $\text{E}_{0}^{(i)}$ are assumed to be shared between schedules. The first model is the complete pooling, thus effectively all model parameters are assumed to be shared.}
\label{tab:LOOIC}
\begin{tabular}{llll}
  \toprule
  	            Model               &  $\text{ED}_{50}^{(i)}$&  $\text{E}_{\text{max}}^{(i)}$ &  LOO-IC   \\        
  	            \midrule
                Model 1            & Shared           & Shared          &   36.0\\ 
                Model 2            & Fixed-effects    & Shared          &   39.8\\ 
                Model 3            & Random-effects   & Shared          &   37.4\\ 
                Model 4            & Fixed-effects    & Fixed-effects   &   41.7\\ 
                Model 5            & Random-effects   & Random-effects  &   41.1\\ 

   \bottomrule
\end{tabular}
\end{table}

\begin{figure}
\centering
\includegraphics[scale=0.5]{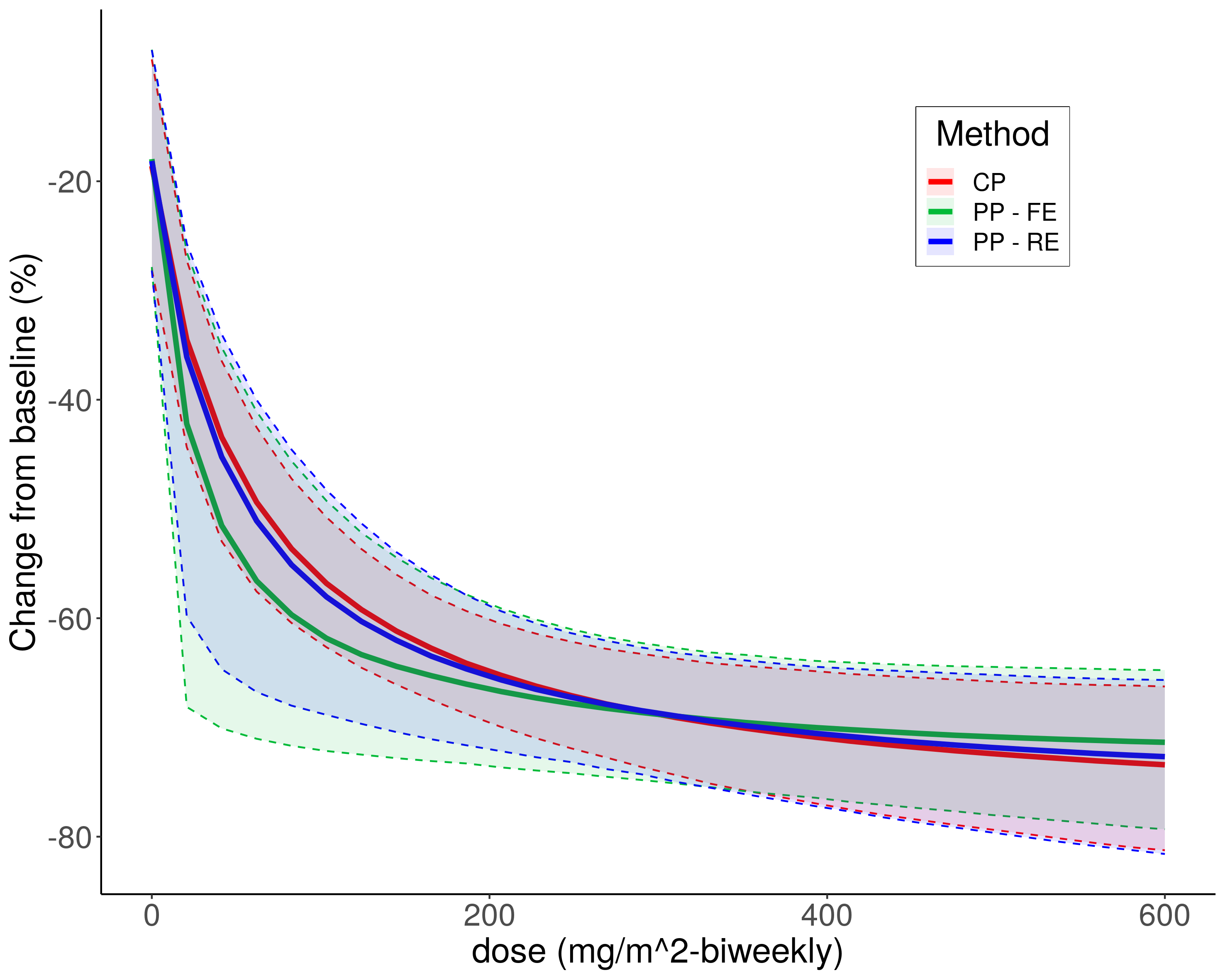}
\caption{Dose-response curve and credible intervals for biweekly schedule obtained by the complete pooling (CP), the partial pooling with fixed-effects (PP - FE), and the partial pooling with random-effects (PP - RE) are shown. See the main text for the descriptions of the methods.\label{fig:Dupi}}
\end{figure}

The posterior estimates obtained by Model 1 (Complete Pooling), Model 2 (PP - FE), and Model 3 (PP - RE) are shown in Table~\ref{tab:DupiRes}. Recall that for Models 2 and 3, the $\text{E}_{\text{max}}$ parameters are shared between schedules. The estimates $\text{ED}_{50}^{\text{weekly}}$ and $\text{ED}_{50}^{\text{monthly}}$ of the complete pooling are calculated by re-scaling the estimate of $\text{ED}_{50}^{\text{biweekly}}$. Across three methods, estimates of $\text{E}_{0}$ are quite similar. For $\text{E}_{\text{max}}$ and $\text{ED}_{50}^{(i)}$, however, the partial pooling with fixed-effects yields different results compared to the complete pooling and the partial pooling with random-effects. The heterogeneity parameter $\tau_{\text{ED}_{50}}$ results in high uncertainty (posterior mean 0.5 with standard deviation of 0.5), indicating the complete pooling is adequate. The estimated dose-response functions $\hat{f}$ by the complete pooling, the partial pooling with fixed-effects, and the partial pooling with random-effects are displayed in Figure~\ref{fig:Dupi}. The $\hat{f}(t)$ are the posterior medians for the dose-response function $f(t)$ evaluated at each $i$ where $i \in \{0, 20.7, \ldots , 600$ (mg/m$^2$-biweekly)$\}$, equidistant sequence between 0 and 600 with 30 elements. Similarly, 95\% equi-tailed credible intervals evaluated at each $i$ are displayed in Figure~\ref{fig:Dupi}. The median dose-response curve obtained by the complete pooling and the partial pooling with random-effects are very similar, which is in alignment with the posterior estimates shown in Table~\ref{tab:DupiRes}. The median dose-response curve estimated by the partial pooling with fixed-effects is slightly different from the complete pooling and the partial pooling with random-effects. As expected, the complete pooling produces the shortest 95\% credible intervals around $\hat{f}$, whereas the partial pooling with fixed-effects gives the widest. Such behaviour was also observed in the simulations. The dupilumab trial is similar to the scenarios when the sample size for each arm is 60, and both $\text{ED}_{50}^{\text{biweekly}}$ and $\text{ED}_{50}^{\text{monthly}}$ do not deviate much from $\text{ED}_{50}^{\text{biweekly}}$, meaning that low heterogeneity in $\text{ED}_{50}^{(i)}$ between schedules. Additionally, Figure~\ref{fig:Dupi_ED50} (Appendix~\ref{app2}) demonstrates the marginal posterior density estimates of $\text{ED}_{50}^{(i)}$ obtained by three methods alongside with the priors used for $\text{ED}_{50}^{(i)}$ in the partial pooling with fixed-effects. The posterior and prior distribution for the $\text{ED}_{50}^{\text{biweekly}}$ parameter are very similar in the partial pooling with fixed-effects. Recall that other than the placebo arm, there is only one arm with weekly schedule, hence indicating the data sparsity problem. In conclusion, although the complete pooling may be sufficient for this particular application, we obtain very similar dose-response estimates by using the partial pooling with random-effects.

\begin{table}[htb]
\centering
\caption{The estimates obtained by analyzing the dupilumab trial. Posterior means and standard deviations obtained by the complete pooling, the partial pooling with fixed-effects (PP - FE), and the partial pooling with random-effects (PP - RE) are shown. See the main text for the descriptions of the methods.}
\label{tab:DupiRes}
\begin{tabular}{lrrrrrrrr}
\toprule
\multicolumn{1}{c}{} &        \multicolumn{2}{c}{CP} & \multicolumn{2}{c}{PP - FE} & \multicolumn{2}{c}{PP - RE}\\
                              \cmidrule(l{3pt}r{3pt}){2-3} \cmidrule(l{3pt}r{3pt}){4-5} \cmidrule(l{3pt}r{3pt}){6-7} 
                                    & Mean  & SD  & Mean & SD & Mean & SD\\
\midrule
$\text{E}_{0}$                      & -18.5     & 4.9     & -18.1   & 5.0   & -18.2  & 5.1   \\
$\text{E}_{\text{max}}$             & -61.0     & 7.4     & -56.9   & 8.0   & -60.0  & 8.6   \\
$\text{ED}_{50}^{\text{weekly}}$    & 32.3      & 15.1    & 20.4    & 27.0  &  30.0  & 29.2  \\
$\text{ED}_{50}^{\text{biweekly}}$ & 64.6      & 30.3    & 37.4    & 35.3  &  56.9  & 40.6  \\
$\text{ED}_{50}^{\text{monthly}}$   & 129.1     & 60.6    & 100.0   & 46.2  &  116.7 & 58.7  \\
$\tau_{\text{ED}_{50}}$             &           &         &         &       &  0.5   & 0.5   \\
\bottomrule
\end{tabular}
\end{table}

%%%%%%%%%%%%%%%%%%%%%%%%%%%%%%%%%%%%%%%%%%%%%%%%%%%%%%
\section{Conclusions and outlook}\label{conclusions}
%%%%%%%%%%%%%%%%%%%%%%%%%%%%%%%%%%%%%%%%%%%%%%%%%%%%%%
An assumption of the homogeneity between schedules can be considered 
unrealistic, hence a partial pooling is more reasonable than the complete pooling. Rather than using schedule-specific fixed-effects in a partial pooling approach, we have proposed to use schedule-specific fixed-effects for the certain parameters such as \(\text{ED}_{50}\), allowing dynamically borrowing information in a fully Bayesian framework. In simulation studies, the proposed method displayed more robust performance in terms of the mean absolute error and coverage probabilities for the dose-response function $f(t)$ compared to the complete pooling. Furthermore, the proposed method produces lower mean absolute error and shorter interval estimates for $f(t)$ across most of the scenarios compared to using schedule-specific fixed-effects in a partial pooling approach.

In this paper, we focused on the Emax model for the dose-response function. To account for the model uncertainty, it is important to consider alternative functions, such as log-linear or exponential. The shrinkage estimation can be applied to such alternative dose-response models, as well. One way of dealing with the model uncertainty is using a model selection criteria (e.g. AIC in the frequentist context) to decide the right functional form. Hence, by using a criteria such as LOO-IC, one can utilize the proposed approach to analyze data from a phase II trial with multiple schedules. Alternatively, a Bayesian model averaging approach \citep{Schorning2016} can be used to deal with uncertainty of dose-response models. Here, we consider phase II trials with multiple schedules. Instead of multiple schedules, one may investigate phase II trials with multiple subgroups, for example multiple patient populations. The proposed method is still applicable for such situations.

The parametrization used in the proposed method, Equation~\eqref{eq:shrinkage1}, can be considered hard to motivate, since an overall mean of schedule-specific estimates does not have a meaningful interpretation. This can be overcome by adopting an asymmetric parametrization of schedule-specific estimates in terms of a reference schedule as follows
\begin{align}
& \text{ED}_{50}^{(k^{*})}  \sim \mathcal{N}(\alpha_{\text{ED50}}, 0)\,\,\,\, (\text{i.e.}\, \text{ED}_{50}^{(k^{*})} = \alpha_{\text{ED50}}) \nonumber\\
& \text{ED}_{50}^{(k)} \sim \mathcal{N}(\alpha_{\text{ED50}}, \beta_{\text{ED50}}^2) \nonumber
\end{align}
where $\alpha_{\text{ED50}}$ and $\beta_{\text{ED50}}$ are the location and scale parameters, respectively \citep{Rover2020}.

Although the partial pooling with random-effects is an improvement to complete pooling and the partial pooling with fixed-effects, the exchangeability assumption bears the risk of too much shrinkage. Perhaps, it is not very desirable to allow borrowing information for the extreme schedule. To overcome this, the exchangeability-nonexchangeability (EXNEX) models \citep{EXNEX} can be considered. EXNEX models can be used to share information across similar schedules, while avoid too much borrowing for the extreme schedule. However, such complicated models should be calibrated well, due to sparse data available in a typical phase II trial.

\clearpage

%=================================================
\section*{Acknowledgements}
%=================================================
We are grateful to Monika Jelizarow and Christian R\"over who contributed valuable comments and pointed us to several important references.

% =======================================
\section*{Conflict of interest}
P.M. is an employee of Galapagos NV. T.F. is a consultant to Galapagos NV.
 MOR106 was jointly discovered by Galapagos NV and MorphoSys, and an MOR106 
 trial was used to motivate and illustrate the investigations presented here.
% =======================================

\clearpage

\appendix

%===========================================================================
\section{How to use the \texttt{ModStan} \textbf{R} package? \label{app1}}
%===========================================================================
The development version of \texttt{ModStan} is available on Github (\url{https://github.com/gunhanb/ModStan}) and can be installed as follows:
\begin{knitrout}
\definecolor{shadecolor}{rgb}{0.969, 0.969, 0.969}\color{fgcolor}
\begin{kframe}
\begin{alltt}
\hlkwd{library}\hlstd{(}\hlstr{"devtools"}\hlstd{)}
\hlkwd{install_github}\hlstd{(}\hlstr{"gunhanb/ModStan"}\hlstd{)}
\end{alltt}
\end{kframe}
\end{knitrout}

The dupilumab trial described in the text is available in the package, and it can be loaded as follows:
\begin{knitrout}
\definecolor{shadecolor}{rgb}{0.969, 0.969, 0.969}\color{fgcolor}
\begin{kframe}
\begin{alltt}
\hlkwd{library}\hlstd{(}\hlstr{"ModStan"}\hlstd{)}
\hlkwd{data}\hlstd{(}\hlstr{"dat.Dupilumab"}\hlstd{)}
\end{alltt}
\end{kframe}
\end{knitrout}
See \texttt{?dat.Dupilumab} for the description of the dataset.

The \texttt{mod\_stan} is the main fitting function of the package. The main computations are executed via the \texttt{rstan} package's \texttt{sampling} function. We can fit the partial pooling method with schedule-specific random-effects for the $\text{ED}_{50}^{(i)}$ parameter as follows:
\begin{knitrout}
\definecolor{shadecolor}{rgb}{0.969, 0.969, 0.969}\color{fgcolor}\begin{kframe}
\begin{alltt}
\hlstd{PP.RE.Dupilumab.stan} \hlkwb{=} \hlkwd{mod_stan}\hlstd{(}\hlkwc{dose} \hlstd{= dose,}
                                \hlkwc{resp} \hlstd{= resp,}
                                \hlkwc{sigma} \hlstd{= sigma,}
                                \hlkwc{schedule} \hlstd{= schedule,}
                                \hlkwc{freq} \hlstd{= freq,}
                                \hlkwc{freq_ref} \hlstd{=} \hlnum{24} \hlopt{*} \hlnum{7} \hlopt{*} \hlnum{8}\hlstd{,}
                                \hlkwc{data} \hlstd{= dat.Dupilumab,}
                                \hlkwc{model} \hlstd{=} \hlstr{"PP-RE"}\hlstd{,}
                                \hlkwc{tau_prior_dist} \hlstd{=} \hlstr{"half-normal"}\hlstd{,}
                                \hlkwc{tau_prior} \hlstd{=} \hlnum{1}\hlstd{,}
                                \hlkwc{chains} \hlstd{=} \hlnum{3}\hlstd{,}
                                \hlkwc{stan_seed} \hlstd{=} \hlnum{111}\hlstd{,}
                                \hlkwc{iter} \hlstd{=} \hlnum{4000}\hlstd{,}
                                \hlkwc{warmup} \hlstd{=} \hlnum{2000}\hlstd{)}
\end{alltt}
\end{kframe}
\end{knitrout}

Convergence diagnostics and the results can be very conveniently obtained using the \texttt{shinystan} package as follows:

\begin{knitrout}
\definecolor{shadecolor}{rgb}{0.969, 0.969, 0.969}\color{fgcolor}
\begin{kframe}
\begin{alltt}
\hlkwd{library}\hlstd{(}\hlstr{"shinystan"}\hlstd{)}
\hlkwd{launch_shinystan}\hlkwd{(as.shinystan}\hlstd{(PP.RE.Dupilumab.stan$fit))}
\end{alltt}
\end{kframe}
\end{knitrout}

The posterior summary statistics can be obtained using the following command:
\begin{knitrout}
\definecolor{shadecolor}{rgb}{0.969, 0.969, 0.969}\color{fgcolor}\begin{kframe}
\begin{alltt}
\hlstd{PP.RE.Dupilumab.stan}
\end{alltt}
\end{kframe}
\end{knitrout}

%==================================================================
\section{Marginal posterior density estimates of $\text{ED}_{50}$ (dupilumab trial) \label{app2}}
\begin{figure}[htb]
\centering
\includegraphics[scale=0.5]{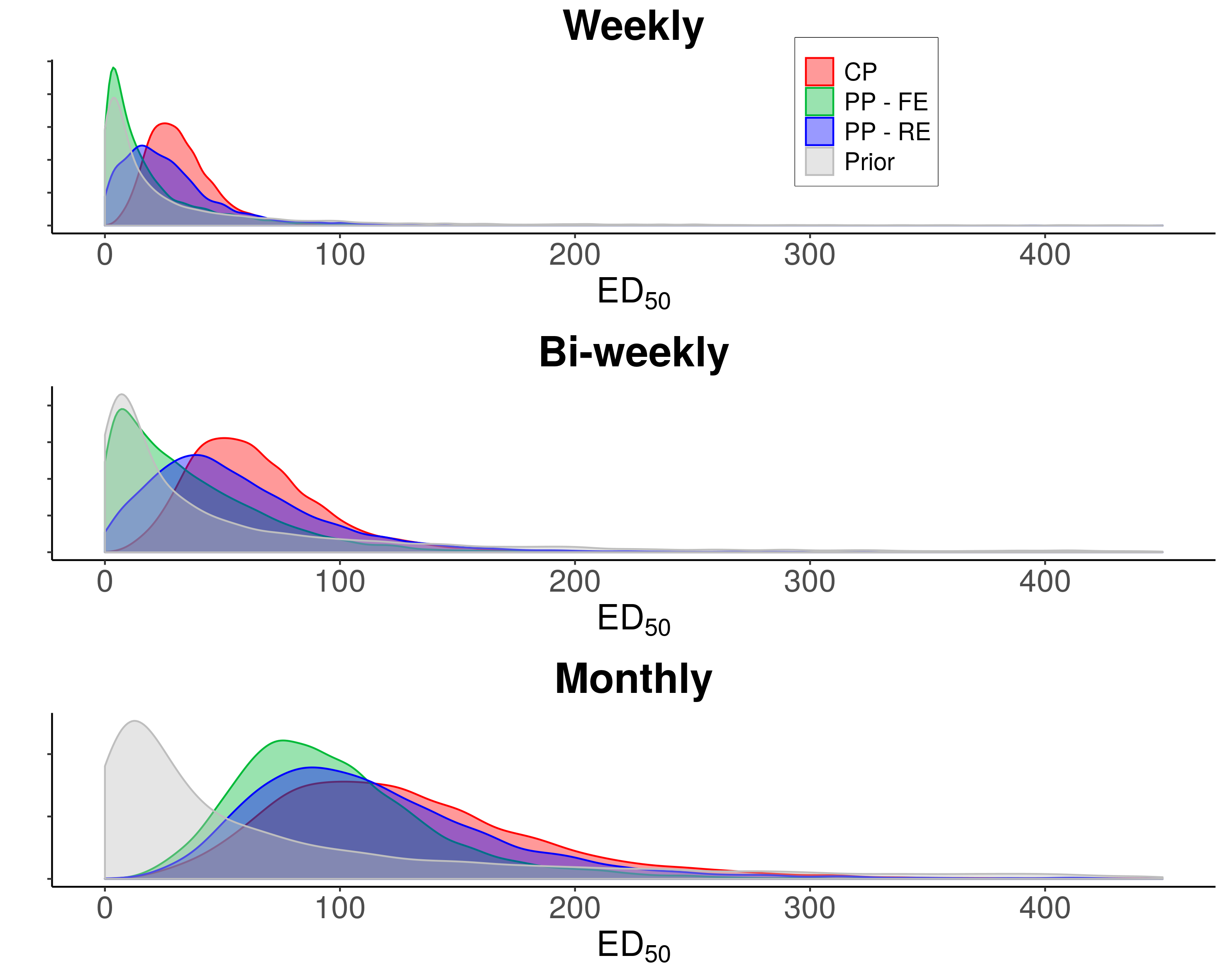}
\caption{Marginal posterior density estimates of $\text{ED}_{50}$ for weekly, biweekly, 
and monthly schedule obtained by the CP, 
the PP - FE, and the PP - RE. Also, shown is prior distributions used
for the PP - FE.\label{fig:Dupi_ED50}}
\end{figure}

%===================================================================
The marginal posterior density estimates obtained by the three methods (CP, PP - FE, PP - RE) are demonstrated
in Figure~\ref{fig:Dupi_ED50}. Also, the prior distribution used
for the PP - FE is shown.

\clearpage

%%%%%%%%%%%%%%%%%%%%%%%%%%%%%%%%%%%%%%%%%%%%%%%%%%%%%%
\bibliography{references}

\begin{thebibliography}{}

\bibitem[Alexander et~al., 2019]{alexander2019}
Alexander, H., Patton, T., Jabbar-Lopez, Z., Manca, A., and Flohr, C. (2019).
\newblock Novel systemic therapies in atopic dermatitis: what do we need to
  fulfill the promise of a treatment revolution?
\newblock {\em F1000Research}, 8(132).

\bibitem[Betancourt and Girolami, 2015]{betancourt2015hamiltonian}
Betancourt, M. and Girolami, M. (2015).
\newblock {\em Current trends in {B}ayesian methodology with applications},
  chapter~4, pages 79--103.
\newblock CRC Press, Boca Raton.

\bibitem[Bornkamp, 2012]{Bornkamp2012}
Bornkamp, B. (2012).
\newblock Functional uniform priors for nonlinear modeling.
\newblock {\em Biometrics}, 68(3):893--901.

\bibitem[Bornkamp, 2014]{Bornkamp2014}
Bornkamp, B. (2014).
\newblock Practical considerations for using functional uniform prior
  distributions for dose-response estimation in clinical trials.
\newblock {\em Biometrical Journal}, 56(6):947--962.

\bibitem[Bornkamp et~al., 2018]{DoseFinding}
Bornkamp, B., Pinheiro, J., and Bretz, F. (2018).
\newblock {\em DoseFinding: Planning and analyzing dose finding experiments.}
\newblock R package version 0.9-16.

\bibitem[Bretz et~al., 2005]{Bretz2005}
Bretz, F., Pinheiro, J., and Branson, M. (2005).
\newblock Combining multiple comparisons and modeling techniques in
  dose-response studies.
\newblock {\em Biometrics}, 61(3):738--748.

\bibitem[Carpenter et~al., 2017]{Stan}
Carpenter, B., Gelman, A., Hoffman, M., Lee, D., Goodrich, B., Betancourt, M.,
  Brubaker, M., Guo, J., Li, P., and Riddell, A. (2017).
\newblock Stan: A probabilistic programming language.
\newblock {\em Journal of Statistical Software}, 76(1):1--32.

\bibitem[Efron and Morris, 1975]{Efron1975}
Efron, B. and Morris, C. (1975).
\newblock Data analysis using {S}tein's estimator and its generalizations.
\newblock {\em Journal of the American Statistical Association},
  70(350):311--319.

\bibitem[Eichenfield and Stein~Gold, 2017]{eichenfield2017}
Eichenfield, L. and Stein~Gold, L. (2017).
\newblock Systemic therapy of atopic dermatitis: Welcome to the revolution.
\newblock {\em Seminars in cutaneous medicine and surgery}, 36(4S):S103--S105.

\bibitem[Feller et~al., 2017]{Feller2017}
Feller, C., Schorning, K., Dette, H., Bermann, G., and Bornkamp, B. (2017).
\newblock Optimal designs for dose response curves with common parameters.
\newblock {\em Annals of Statistics}, 45(5):2102--2132.

\bibitem[Freidlin and Korn, 2013]{Freidlin1326}
Freidlin, B. and Korn, E. (2013).
\newblock Borrowing information across subgroups in phase {II} trials: Is it
  useful?
\newblock {\em Clinical Cancer Research}, 19(6):1326--1334.

\bibitem[Friede et~al., 2017]{friede2017}
Friede, T., R\"over, C., Wandel, S., and Neuenschwander, B. (2017).
\newblock Meta-analysis of few small studies in orphan diseases.
\newblock {\em Research Synthesis Methods}, 8(1):79--91.

\bibitem[Gelman, 2006]{gelman2006}
Gelman, A. (2006).
\newblock Prior distributions for variance parameters in hierarchical models
  (comment on article by browne and draper).
\newblock {\em Bayesian Analysis}, 1(3):515--534.

\bibitem[Giugliano et~al., 2012]{Giugliano2012}
Giugliano, R., Desai, N., Kohli, P., Rogers, W., Somaratne, R., Huang, F., Liu,
  T., Mohanavelu, S., Hoffman, E., McDonald, S., Abrahamsen, T., Wasserman, S.,
  Scott, R., and Sabatine, M. (2012).
\newblock Efficacy, safety, and tolerability of a monoclonal antibody to
  proprotein convertase subtilisin/kexin type 9 in combination with a statin in
  patients with hypercholesterolaemia ({LAPLACE-TIMI 57}): a randomised,
  placebo-controlled, dose-ranging, phase 2 study.
\newblock {\em The Lancet}, 380(9858):2007--2017.

\bibitem[Greenland, 2000]{Greenland2000}
Greenland, S. (2000).
\newblock {Principles of multilevel modelling.}
\newblock {\em International Journal of Epidemiology}, 29(1):158--167.

\bibitem[G\"unhan et~al., 2020]{Guenhan2020Meta}
G\"unhan, B., R\"over, C., and Friede, T. (2020).
\newblock Random-effects meta-analysis of few studies involving rare events.
\newblock {\em Research Synthesis Methods}, 11(1):74--90.

\bibitem[Jones et~al., 2011]{Jones2011}
Jones, H., Ohlssen, D., Neuenschwander, B., Racine, A., and Branson, M. (2011).
\newblock Bayesian models for subgroup analysis in clinical trials.
\newblock {\em Clinical Trials}, 8(2):129--143.

\bibitem[{Mayo Clinic}, 2018]{atopic}
{Mayo Clinic} (2018).
\newblock {Atopic Dermatitis}.
\newblock
  \url{https://www.mayoclinic.org/diseases-conditions/atopic-dermatitis-eczema/symptoms-causes/syc-20353273}.
\newblock Updated March, 2018. Accessed January, 2020.

\bibitem[M\"ollenhoff et~al., 2019]{Mollenhof2019}
M\"ollenhoff, K., Bretz, F., and Dette, H. (2019).
\newblock Equivalence of regression curves sharing common parameters.
\newblock {\em Biometrics}, pages 1--12.

\bibitem[{MorphoSys AG}, 2019]{MOR106}
{MorphoSys AG} (2019).
\newblock {MOR106} clinical development in atopic dermatitis stopped.
\newblock
  \url{https://www.morphosys.com/media-investors/media-center/morphosys-ag-mor106-clinical-development-in-atopic-dermatitis-stopped}.
\newblock Updated October, 2019. Accessed January, 2020.

\bibitem[Neuenschwander et~al., 2016]{EXNEX}
Neuenschwander, B., Wandel, S., Roychoudhury, S., and Bailey, S. (2016).
\newblock Robust exchangeability designs for early phase clinical trials with
  multiple strata.
\newblock {\em Pharmaceutical Statistics}, 15(2):123--134.

\bibitem[R\"over and Friede, 2020]{Rover2020}
R\"over, C. and Friede, T. (2020).
\newblock Dynamically borrowing strength from another study through shrinkage
  estimation.
\newblock {\em Statistical Methods in Medical Research}, 29(1):293--308.

\bibitem[Ruberg, 1995]{ruberg}
Ruberg, S. (1995).
\newblock Dose response studies i. some design considerations.
\newblock {\em Journal of Biopharmaceutical Statistics}, 5(1):1--14.

\bibitem[Schorning et~al., 2016]{Schorning2016}
Schorning, K., Bornkamp, B., Bretz, F., and Dette, H. (2016).
\newblock Model selection versus model averaging in dose finding studies.
\newblock {\em Statistics in Medicine}, 35(22):4021--4040.

\bibitem[Spiegelhalter et~al., 2004]{spiegelhalter2004}
Spiegelhalter, D., Abrams, K., and Myles, J. P.~d. (2004).
\newblock {\em Bayesian Approaches to clinical trials and health-care
  evaluation.}
\newblock West Sussex: CRC Press.

\bibitem[Tha{\c{c}}i et~al., 2016]{dupilumab}
Tha{\c{c}}i, D., Simpson, E., Beck, L., Bieber, T., Blauvelt, A., Papp, K.,
  Soong, W., Worm, M., Szepietowski, J., Sofen, H., et~al. (2016).
\newblock Efficacy and safety of dupilumab in adults with moderate-to-severe
  atopic dermatitis inadequately controlled by topical treatments: a
  randomised, placebo-controlled, dose-ranging phase 2b trial.
\newblock {\em Lancet}, 387(10013):40--52.

\bibitem[Thomas et~al., 2014]{Thomas2014}
Thomas, N., Sweeney, K., and Somayaji, V. (2014).
\newblock Meta-analysis of clinical dose-response in a large drug development
  portfolio.
\newblock {\em Statistics in Biopharmaceutical Research}, 6(4):302--317.

\bibitem[Varadhan, 2015]{alabama}
Varadhan, R. (2015).
\newblock {\em alabama: Constrained nonlinear optimization.}
\newblock R package version 2015.3-1.

\bibitem[Vehtari et~al., 2017]{Vehtari2017}
Vehtari, A., Gelman, A., and Gabry, J. (2017).
\newblock Practical {B}ayesian model evaluation using leave-one-out
  cross-validation and {WAIC}.
\newblock {\em Statistics and Computing}, 27(5):1413--1432.

\bibitem[Viele et~al., 2014]{viele}
Viele, K., Berry, S., Neuenschwander, B., Amzal, B., Chen, F., Enas, N., Hobbs,
  B., Ibrahim, J.~G., Kinnersley, N., Lindborg, S., Micallef, S., Roychoudhury,
  S., and Thompson, L. (2014).
\newblock Use of historical control data for assessing treatment effects in
  clinical trials.
\newblock {\em Pharmaceutical Statistics}, 13(1):41--54.

\end{thebibliography}
%%%%%%%%%%%%%%%%%%%%%%%%%%%%%%%%%%%%%%%%%%%%%%%%%%%%%%

\end{document}